\documentclass[iop]{emulateapj}
\begin{document}
\title{Observational Signatures of Tilted Black Hole Accretion Disks from Simulations}
\shorttitle{Signatures of Tilted Accretion Disks}
\shortauthors{Dexter \& Fragile}
\author{Jason Dexter}
\affil{Department of Physics, University of Washington, Seattle, WA 98195-1560, USA}
\email{jdexter@u.washington.edu}
\author{P. Chris Fragile}
\affil{Department of Physics \& Astronomy, College of Charleston, Charleston, SC 29424, USA}
\begin{abstract}
Geometrically thick accretion flows may be present in black hole X-ray binaries observed in the low/hard state and in low-luminosity active galactic nuclei. Unlike in geometrically thin disks, the angular momentum axis in these sources is not expected to align with the black hole spin axis. We compute images from three-dimensional general relativistic magnetohydrodynamic simulations of misaligned (tilted) accretion flows using relativistic radiative transfer, and compare the estimated locations of the radiation edge with expectations from their aligned (untilted) counterparts. The radiation edge in the tilted simulations is independent of black hole spin for a tilt of $15^\circ$, in stark contrast to the results for untilted simulations, which agree with the monotonic dependence on spin expected from thin accretion disk theory. Synthetic emission line profiles from the tilted simulations depend strongly on the observer's azimuth, and exhibit unique features such as broad ``blue wings." Coupled with precession, the azimuthal variation could generate time fluctuations in observed emission lines, which would be a clear ``signature'' of a tilted accretion flow. Finally, we evaluate the possibility that the observed low- and high-frequency quasi-periodic oscillations (QPOs) from black hole binaries could be produced by misaligned accretion flows. Although low-frequency QPOs from precessing, tilted disks remains a viable option, we find little evidence for significant power in our light curves in the frequency range of high-frequency QPOs.
\end{abstract}
\maketitle

\section{Introduction}

In standard thin disk accretion theory \citep{shaksun1973,novthorne}, the angular momentum axis of the accretion flow is assumed to be aligned with the black hole spin axis. \citet{bardeenpetterson1975} found that even if the initial angular momentum axis of the accretion flow is misaligned from the black hole spin axis, the inner part of the disk will still align on the viscous timescale. However, this so-called ``viscous" regime only operates when $H/R \lesssim \alpha$, where $H/R$ is the scale height of the accretion disk, and $\alpha$ is the parameterized viscosity \citep{papaloizoulin1995}. This is applicable in active galactic nuclei (AGN) and the high/soft or thermal state of black hole X-ray binaries. On the other hand, advection-dominated accretion flows (ADAFs) are expected in the low/hard state of black hole X-ray binaries \citep{narayanyi1995adaf,esinetal1997} and in low-luminosity AGN. ADAFs are unable to cool through efficient radiation, and are geometrically thick. It is likely that the accretion flow in many of these sources is misaligned, or ``tilted.''

Contemporary general relativistic MHD simulations \citep[GRMHD, ][]{devilliers2003,gammie2003} currently provide the most physically realistic description of the inner portion of accretion flows around spinning black holes. Radiation can be calculated from these simulations in post-processing by assuming that it is dynamically and thermodynamically negligible. This method has been used to look for high frequency quasi-periodic oscillations (HFQPOs) in simulated data \citep{schnittman2006} and to create radiative models of Sagittarius A* \citep{noble2007,moscibrodzka2009,dexter2009,dexteretal2010}. 

All of this work assumed alignment between the angular momentum axis of the accretion flow and the black hole spin axis. \citet{fragile2007,fragileetal2009,fragiletilt2009} were the first to do GRMHD simulations of disks with a tilt between these two axes. These new simulations yielded a number of unexpected features. First, the main body of the disk remained tilted with respect to the symmetry plane of the black hole; thus there was no indication of a Bardeen-Petterson effect in the disk at large. The torque of the black hole instead principally caused a global precession of the main disk body \citep{fragile2005,fragile2007}. The time-steady structure of the disk was also warped, with latitude-dependent radial epicyclic motion driven by pressure gradients attributable to the warp \citep{fragile2008}. The tilted disks also truncated at a larger radius than expected for an untilted disk. In fact, based on dynamical measures, the inner edge of these tilted disks was found to be independent of black hole spin \citep{fragiletilt2009}, in sharp contrast to the expectation that accretion flows truncate at the marginally stable orbit of the black hole. Finally, \citet{henisey2009} found evidence for trapped inertial waves in a simulation with a black spin $a=0.9$, producing excess power at a frequency $118 (M/10 M_\sun)^{-1}$ Hz.

In this work we use relativistic ray tracing to produce images and light curves of some of these numerically simulated tilted and untilted black-hole accretion disks. Our goal in this paper is to discuss observable differences between the two types of accretion flows, and to identify observational signatures of tilted black hole accretion disks.

\begin{figure}
\epsscale{1.2}
\plotone{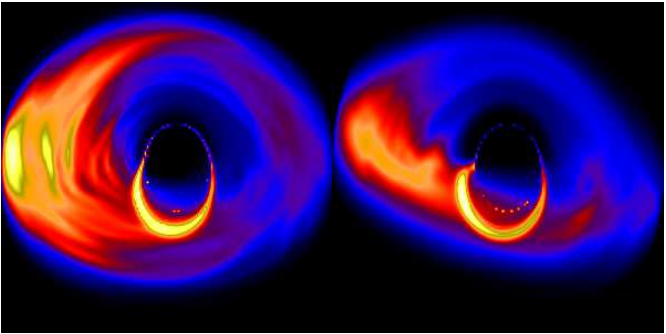}
\caption{\label{imgs}Sample images of the thermal emission model for the 90h (left) and 915h (right) simulations at $60^\circ$ inclination. The observed photon energy is $E_0=10$keV for a $10 M_\sun$ black hole, and each panel is $54 M$ across. The color scale is linear, increasing from blue to red to yellow to white.}
\end{figure}

\section{Methods}

\subsection{Simulation Data}

The simulations used here are from \citet{fragile2007,fragileetal2009,fragiletilt2009}. The parameters are given in Table \ref{sims}. All of the simulations used the Cosmos++ GRMHD code \citep{anninos2005}, with an effective resolution of $128^3$ for the spherical-polar grid (except near the poles where the grid was purposefully underresolved) and $128\times64\times64\times6$ for the cubed-sphere grid. The simulations were initialized with an analytically solvable, time-steady, axisymmetric gas torus \citep{devilliers2003}, threaded with a weak, purely poloidal magnetic field that follows the isodensity contours and has a minimum $P_{gas}/P_{mag}=10$ initially. The magnetorotational instability (MRI) arose naturally from the initial conditions, and the disk quickly became fully turbulent. The simulations were all evolved for $\sim$8000M, or $\sim$40 orbits at $r=10$M in units with $G=c=1$. Only data from the final $2/3$ of the simulation are used in this analysis, once the disks are fully turbulent as measured by a peak in the accretion rate and in the mass inside of $r=10$M. This is chosen to utilize as much of the simulation data as possible, and none of our results depend on which time interval 
in the simulation is used.

\begin{deluxetable}{cccc}
\tabletypesize{\scriptsize}
\tablecaption{Simulation Parameters \label{sims}}
\tablewidth{0pt}
\tablehead{
\colhead{Simulation} & \colhead{$a/M$} & \colhead{Tilt} &
\colhead{Grid} \\
\colhead{} & \colhead{} & \colhead{Angle} & 
\colhead{} 
}
\startdata
0H\tablenotemark{a} & 0 & ... & Spherical-polar \\
315H\tablenotemark{b} & 0.3 & $15^\circ$ & Spherical-polar \\
50H\tablenotemark{a} & 0.5 & $0^\circ$ & Cubed-sphere \\
515H\tablenotemark{a} & 0.5 & $15^\circ$ & Spherical-polar \\
715H\tablenotemark{b} & 0.7 & $15^\circ$ & Spherical-polar \\
90H\tablenotemark{c} & 0.9 & $0^\circ$ & Spherical-polar \\
915H\tablenotemark{c} & 0.9 & $15^\circ$ & Spherical-polar
\enddata
\tablenotetext{a}{\citet{fragileetal2009}}
\tablenotetext{b}{\citet{fragiletilt2009}}
\tablenotetext{c}{\citet{fragile2007}}

\end{deluxetable}

These simulations all evolved an internal energy equation, and injected entropy at shocks. Such a formulation does not conserve energy, and produces a more slender, cooler torus than conservative formulations which capture the heat from numerical reconnection of magnetic fields \citep{fragile2009}. The scale height spanned the range $H/R \sim 0.05-0.1$ in these simulations, with larger scale heights for higher spin simulations.

\subsection{Ray Tracing}

Relativistic radiative transfer is computed from simulation data via ray tracing. Starting from an observer's camera, rays are traced backwards in time assuming they are null geodesics (geometric optics approximation), using the public code described in \citet{dexteragol2009}. In the region where rays intersect the accretion flow, the radiative transfer equation is solved along the geodesic \citep{broderick2006} in the form given in \citet{fuerstwu2004}, which then represents a pixel of the image. This procedure is repeated for many rays to produce an image, and at many time steps of the simulation to produce time-dependent images (movies). Light curves are computed by integrating over the individual images. Sample images of two simulations are given in Figure \ref{imgs}. Doppler beaming causes asymmetry in the intensity from approaching (left) and receding (right) fluid. Photons emitted from the far side of the accretion flow are deflected toward the observer, causing it to appear above the black hole. The thick, central ring is due to gravitational lensing from material passing under the black hole, while the underresolved circular ring is caused by photons that orbit the black hole one or more times before escaping. These ring features are in excellent agreement with the predictions made by \citet{viergutz1993}.

To calculate fluid properties at each point on a ray, the spacetime coordinates of the geodesic are transformed from Boyer-Lindquist to the Kerr-Schild coordinates used in the simulation. Since the accretion flow is dynamic, light travel time delays along the geodesic are taken into account. Data from the sixteen nearest zone centers (eight on the simulation grid over two time steps) are interpolated to each point on the geodesic. Between levels of resolution near the poles on the spherical-polar grid, data from the higher resolution layer are averaged to create synthetic lower resolution points, which are then interpolated. Very little emission originates in the underresolved regions of the simulation.

The simulations provide mass density, pressure, velocity and magnetic field in code units. These are converted into cgs units following the procedure described in \citet{schnittman2006} and \citet{dexter2009}. The length- and time-scales are set by the black hole mass, taken to be $10 M_{\sun}$ throughout.

We consider two emission models. The thin line emissivity from \citet{schnittman2006} is a toy model that traces the mass density in the accretion flow. The thermal emission model from \citet{schnittman2006} uses free-free emission and absorption coefficients, and is used as a model for the high/soft state. Although we do not expect tilted disks to accurately represent the high/soft state, this model may be appropriate for sources radiating at an appreciable fraction of Eddington, where the infall time is shorter than the radiative diffusion time and the accretion flow becomes geometrically ``slim." When taking the temperature from the ideal gas law rather than the radiation-dominated equation of state used in \citet{schnittman2006}, this model may be qualitatively appropriate for modeling the low/hard state in X-ray binaries or low-luminosity AGN.

In \S \ref{radedge}, we consider emission from inside of $r=15$M, while in \S \ref{lines} and \ref{var} fluid inside of $r=25$M is used for the ray tracing. For all results here, we take the temperature from the ideal gas law rather than assuming a radiation-dominated equation of state. 
All of our results are qualitatively identical when using the radiation-dominated equation of state to calculate the temperature.

\begin{figure}
\epsscale{1.2}
\plotone{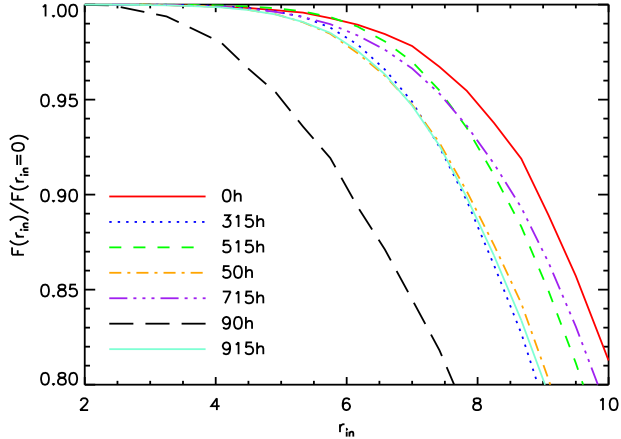}
\caption{\label{tfluxrline1}Comparison of relative intensities for all simulations using the thin line emissivity. The flux from grids of images over observer time, inclination and azimuth for each simulation have been averaged to create these curves.}
\end{figure}

\begin{figure}
\epsscale{1.2}
\plotone{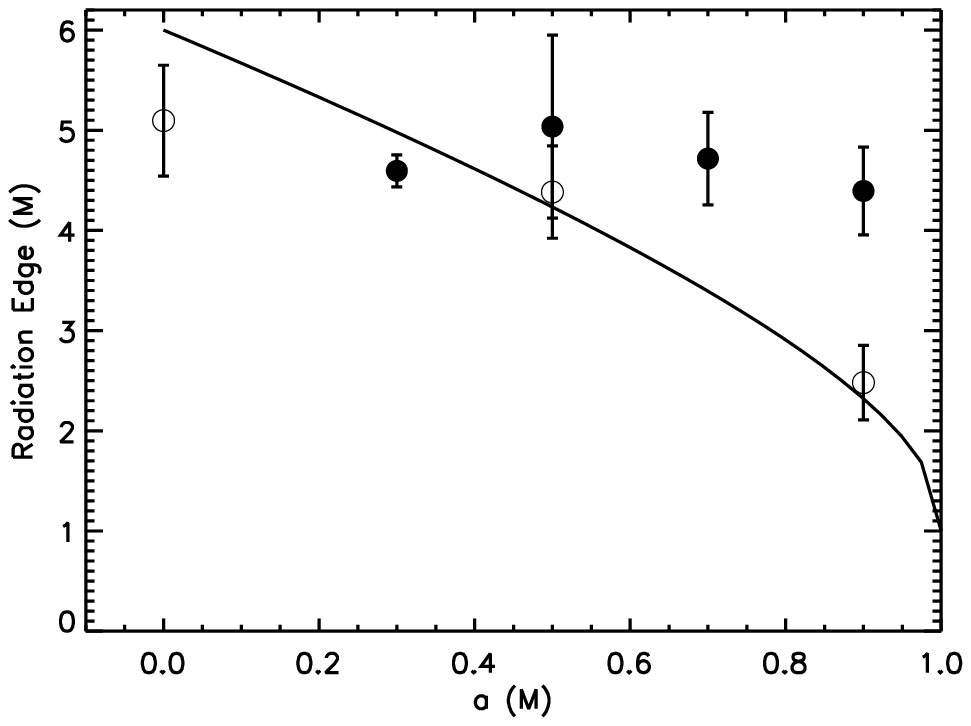}
\caption{\label{rinline1}Radiation edge as a function of spin for untilted (open) and tilted (solid) simulations for the thin line emissivity. The error bars show the one standard deviation time variability in the radiation edge, averaged over other parameters. The solid line is the marginally stable orbit.}
\end{figure}

\begin{figure}
\epsscale{1.2}
\plotone{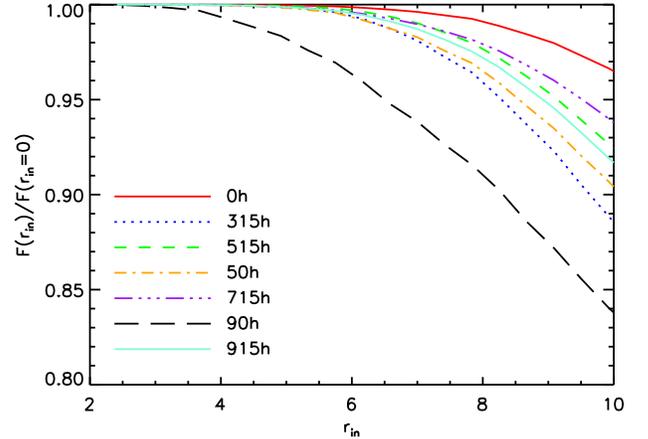}
\caption{\label{tfluxrline2} Comparison of relative intensities for all simulations using the thermal emissivity at $E_0=1$ keV. The flux from grids of images over observer time, inclination and azimuth for each simulation have been averaged to create these curves.}
\end{figure}

\begin{figure}
\epsscale{1.2}
\plotone{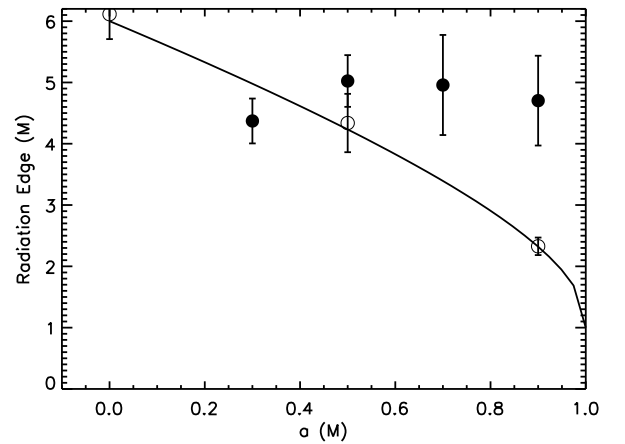}
\caption{\label{rinline2}Radiation edge as a function of spin for untilted (open) and tilted (solid) simulations for the thermal emissivity at $E_0=1$keV. The error bars show the one standard deviation time variability in the radiation edge, averaged over other parameters. The solid line is the marginally stable orbit.}
\end{figure}

\section{Results}

\subsection{Radiation Edge}
\label{radedge}
Inferring the inner edge of accretion flows is important for attempts to measure spin from broad iron lines \citep[e.g.][]{wilms2001} or continuum fitting \citep[e.g.][]{shafee2006,davis2006}. Such measurements assume that the disk has a sharp cutoff at the innermost stable circular orbit, which depends on spin \citep{bardeen1972}. \citet{fragiletilt2009} used four different dynamical measures from \citet{krolikhawley2002} to compare the inner edges of simulated tilted and untilted accretion disks. Here we use the ray traced models to locate the ``radiation edge," the radius inside of which the contribution to the total flux is negligible. 

For each emission model, images are calculated for all simulations over a grid of observer inclination, observer time, and observer azimuth (for the tilted simulations). We then compute images cutting out fluid inside of successive values of the radius $r_{\mathrm{in}}$. The radiation edge is functionally defined as the radius where the ratio of intensities, $F(r_{\mathrm{in}})/F(0)$, drops below an arbitrary fraction $f$, chosen so that the untilted radiation edge agrees as well as possible with $r_{\mathrm{ms}}$, the marginally stable orbit.

Figure \ref{tfluxrline1} shows a plot of $F(r_{\mathrm{in}})/F(0)$ as a function of $r_{\mathrm{in}}$ averaged over observer time, azimuth and inclination for the thin line emissivity. From these curves we extract values of the radiation edge, $r_{\mathrm{edge}}$. Results are shown in Figure \ref{rinline1}, where the error bars are computed from the standard deviation of $r_{\mathrm{edge}}$ as a function of time, averaged over the other parameters. This result agrees well with the dynamical measures from \citet{fragiletilt2009}. While the radiation edge moves in towards the black hole with increasing spin for untilted simulations, there is no such trend in the tilted simulations. Instead, the radiation edge appears to be independent of spin.

Figures \ref{tfluxrline2} and \ref{rinline2} show the same plots for the thermal emission model with observed photon energy $E_0=1$ keV. The conclusions are identical with this emission model. The untilted simulations have radiation edges which agree quite well with $r_{\mathrm{ms}}$, while the tilted simulations show no correlation between spin and $r_{\mathrm{edge}}$. Again, these results are consistent with \citet{fragiletilt2009}, although we find no trend of \emph{increasing} radiation edge with spin, as was found for a couple of the dynamical measures used in \citet{fragiletilt2009}. Plots from other observed photon energies are not shown; although the relative flux falls off much more quickly with increasing $r_{\mathrm{in}}$ at higher photon energies, the results for the radiation edge remain completely unchanged. 

\begin{figure*}
\epsscale{1.2}
\plotone{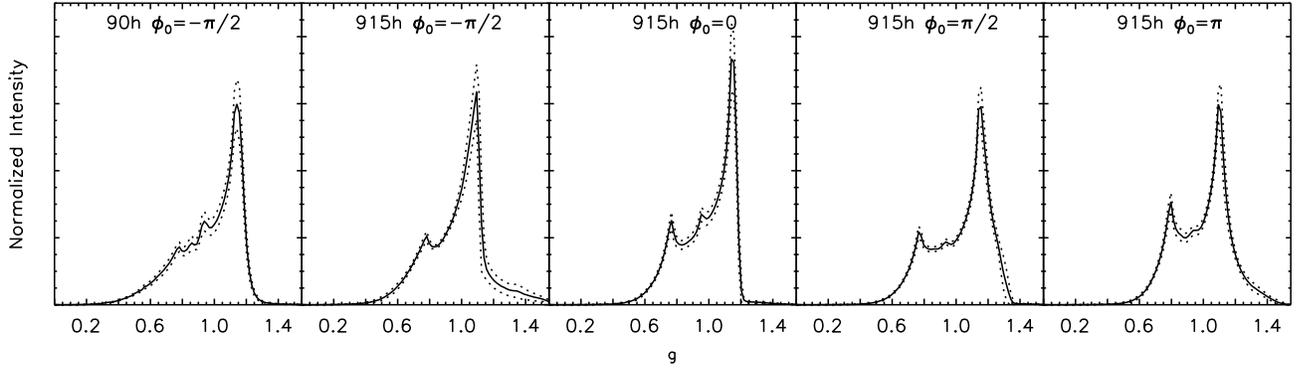}
\caption{\label{iline}Emission line profiles for simulations with a=0.9. The emissivity is $j \propto \rho r^{-3}$ and the observer inclination is $60^\circ$ in all cases. The dotted lines show the $1\sigma$ range, taken from the time variability.}
\end{figure*}

\begin{figure}
\epsscale{1.2}
\plotone{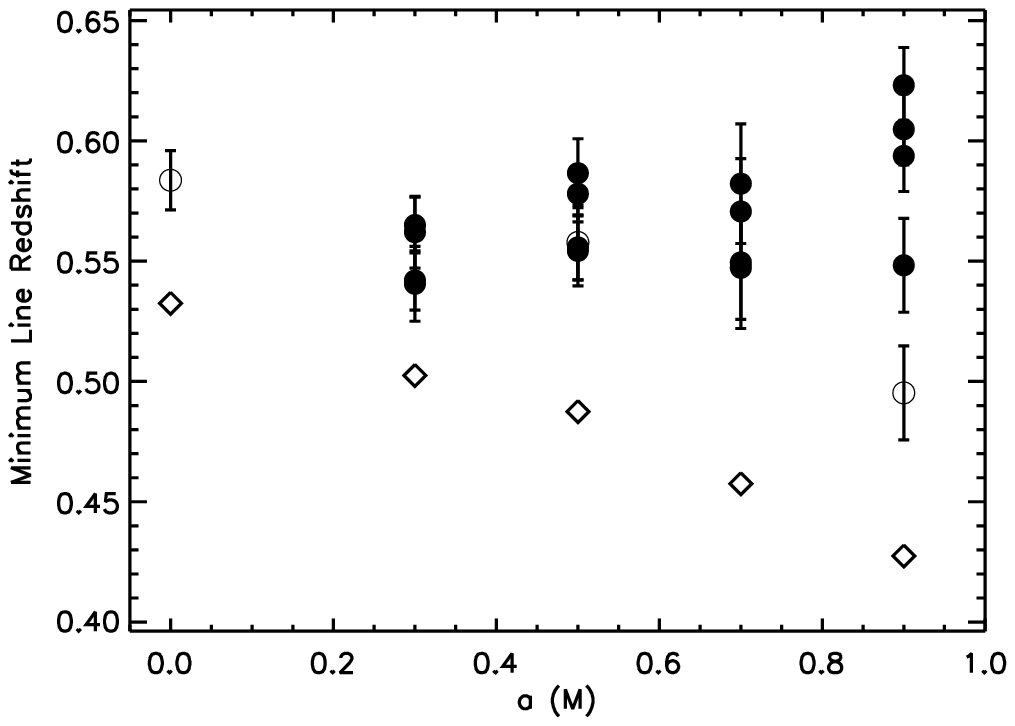}
\caption{\label{ilinew1}Minimum line energy vs. spin for all simulations. The tilted (untilted) simulations are denoted by solid (open) circles, and four observer azimuths are plotted for the tilted simulations. The open diamonds are from a thin disk in the equatorial plane with an emissivity $j \propto r^{-3}$ similar to that used in \citet{schnittmanbertschinger2004,dexteragol2009}. The minimum line energy is defined as the lowest energy contained in the set of intensities comprising 99\% of the total line intensity. The $1\sigma$ errors are taken from the time variability.}
\end{figure}

\begin{figure}
\epsscale{1.2}
\plotone{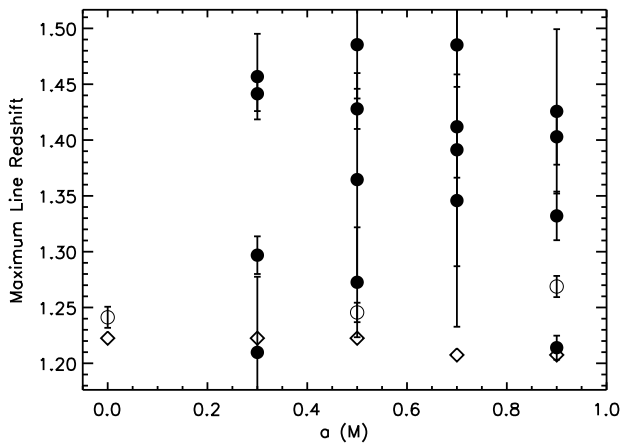}
\caption{\label{ilinew2}As in Figure \ref{ilinew1}, but for the maximum line energy.}
\end{figure}

\subsection{Emission Line Profiles}
\label{lines}
Spectra from AGN and X-ray binaries typically include strong emission and absorption features. As the observed line shapes are sensitive to both the velocity of the emitting/absorbing fluid, and also to the local gravitational redshift, they can provide information about the dynamics of the accretion flow \citep{fabianetal1989,laor1991}. 

Untilted accretion flows have nearly Keplerian velocity distributions outside the marginally stable orbit, where the velocities smoothly transition to plunging. Simulated tilted accretion disks, on the other hand, show three major differences. The Keplerian velocity structure is now tilted. 

Secondly, the warped structure of the tilted disks leads to epicyclic motions with velocity magnitudes comparable to the local geodesic orbital velocity \citep{fragile2008}. Finally, the larger radiation edge values of the tilted disks identified in \S \ref{radedge} means that the transition to plunging orbits occurs at larger radius than in untilted disks.

These effects indicate that we should expect a number of differences in line profiles from tilted accretion flows \citep{fragileetal2005}. The maximum blueshift should be larger for tilted accretion disks, except for edge-on viewing. For $i < 90^\circ - \beta$, where $i$ is the observer's inclination angle and $\beta$ is the initial tilt angle, both relative to the black hole spin axis, the tilted accretion flow should mimic an untilted one with a larger inclination. In contrast, the red wing should be less pronounced in the tilted disks due to their larger truncation radii. On the redshifted side, tilted disks behave similar to lower spin, untilted disks. 

Producing a detailed reflection spectrum would require a significant number of assumptions to model the metallicity, ionization levels, and incident X-ray flux throughout the accretion flow. For simplicity, we instead use toy model emissivities of the form $j \propto \rho r^{-s}$, where $\rho$ is the fluid mass density, $j$ is the photon-energy integrated emissivity and $s=2$,$3$. The two values correspond to assuming the emitted line flux is proportional to the incident flux from an irradiating source on the spin axis and to local dissipation of heat, respectively \citep[e.g.,][]{fragileetal2005}. This simple form allows us to focus on general features to be expected from emission lines from tilted black hole accretion disks.

Figure \ref{iline} shows sample line profiles for an inclination of $i=60^\circ$ for four observer azimuths from the 915h simulation. Only a single observer azimuth from the 90h simulation is shown, since the time-averaged emission line is independent of observer azimuth for untilted simulations. In all cases, the lines consist of a strong peak near the rest energy of the line ($g \equiv E_0 / E_{em} = 1$), a smaller peak at lower energy and a ``red wing," whose extent and strength depends on the amount of emission arising very close to the black hole (small $g$). The location of the ``blue" peak (large $g$) depends on the maximum velocity along the line of sight in the accretion flow. For an untilted disk, this corresponds directly to the observer's inclination angle, since all fluid velocities are essentially in the equatorial plane. 

For the tilted model shown in Figure \ref{iline}, the location and strength of the blue peak changes significantly with observer azimuth. When the angular momentum axis of the accretion flow is in the plane of the sky ($-\pi/2 \lesssim \phi_0 \lesssim -\pi/4$, depending on the simulation time), 
its fluid velocities are maximally aligned with the observer's line of sight, leading to the largest blueshifts. This is the same condition as an untilted disk being viewed edge-on. For other orientations, the blue tail can extend to significantly higher photon energies in the tilted simulations because the largest effective inclination is approximately $i_{\mathrm{eff}} = i + \beta$. When the accretion flow is not edge-on, there will exist orientations where $i_{\mathrm{eff}} > i$, and the blue peak for a tilted simulation will occur at higher energy than possible for untilted accretion flows. The red wing, on the other hand, remains largely unchanged with observer azimuth, since it is caused by gravitational redshifts rather than Doppler boosts. Since the radiation edge for the 915h simulation was found to occur at significantly larger radius than that of 90h, it is expected that the red wing should extend further in the 90h simulation. The effect is subtle, but identifiable in Figure \ref{iline}.

To quantify these trends, for all simulations we compute the extent of the line profile, as well as the strengths and locations of their red and blue peaks. Most clear are the results for the line extents, shown for $i=60^\circ$ in Figures \ref{ilinew1} and \ref{ilinew2}. As expected, the red wing extends to lower photon energies at higher spins for untilted simulations, while there is no similar trend for the tilted models. Also as expected, the blue wing extends to systematically higher photon energies in the tilted simulations because of the difference between $i_{\mathrm{eff}}$ and $i$ noted above and the epicyclic motion in the tilted simulations.

Perhaps the most striking feature of the line profiles is the variation with observer azimuth seen in all tilted simulations. These changes in line shape between different observer azimuths are typically larger than the full range of changes seen between different spins for untilted simulations. This suggests that the most powerful means of recognizing a tilted accretion disk may be to measure changes in an emission line profile over time as the disk precesses.

\begin{figure*}
\epsscale{1.2}
\plotone{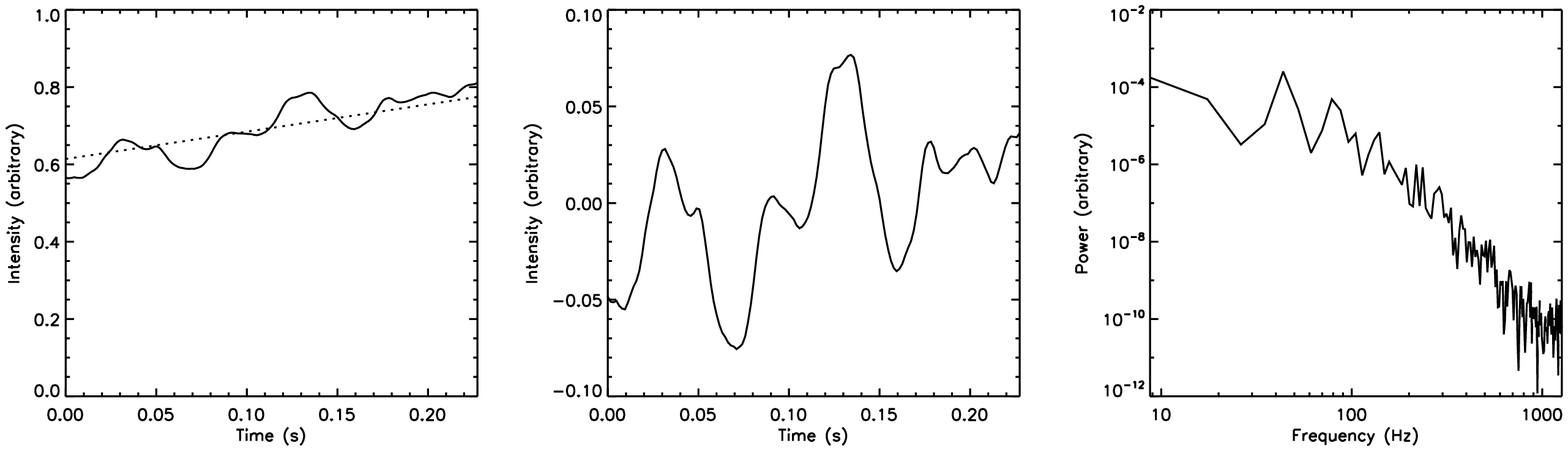}
\caption{\label{lexample}Sample light curve and linear fit (left), light curve with linear fit subtracted (middle) and power spectrum (right). The units are scaled to a 10 solar mass black hole.}
\end{figure*}

\begin{figure*}
\epsscale{1.2}
\plotone{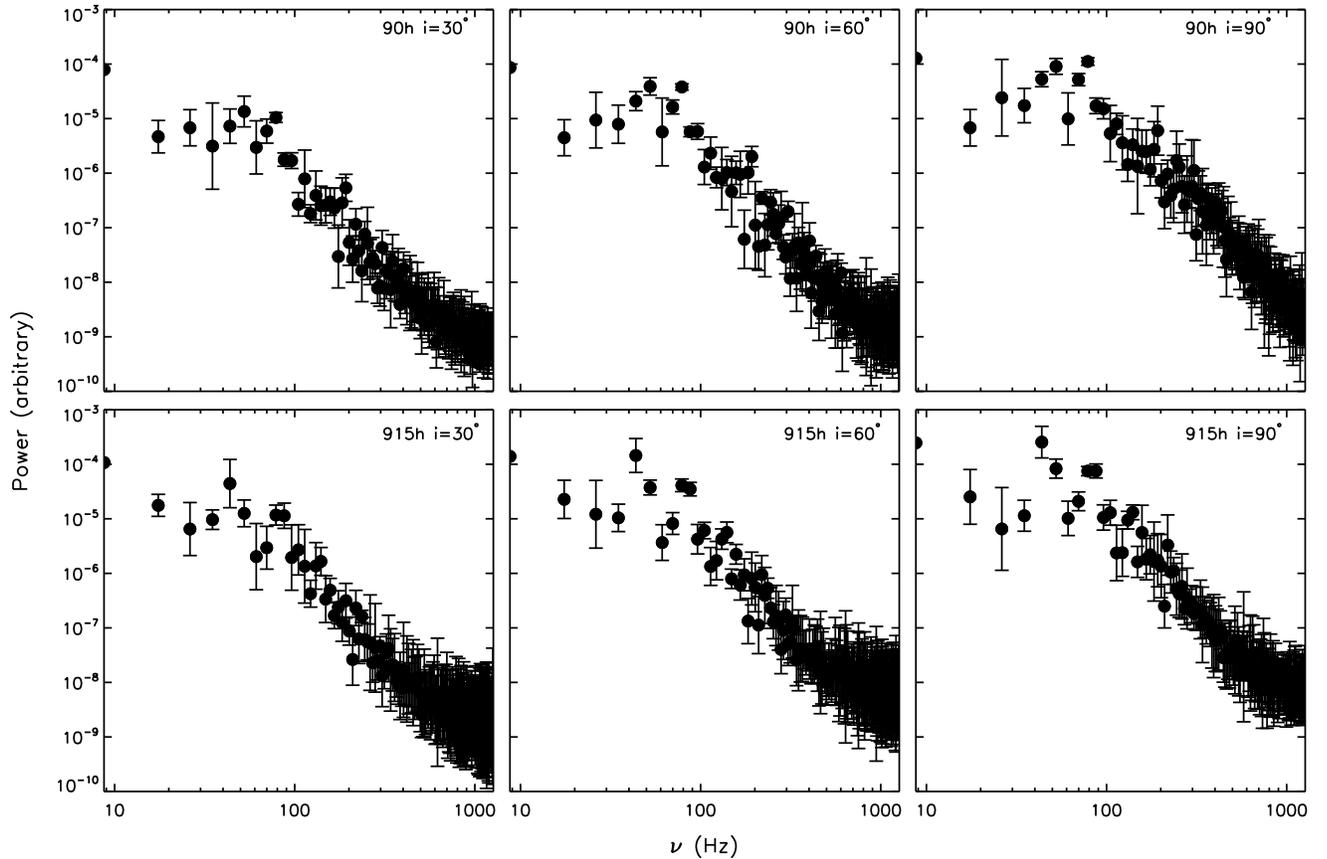}
\caption{\label{allpsds}Median power spectra for $i=30^\circ$, $60^\circ$, $90^\circ$ from the 90h and 915h simulations. The errors are estimated from the standard deviations of the set of power spectra at observed photon energies of $1$, $3$, $10$ keV at four observer azimuths. All power spectra are well described by the broken power law model, with break frequencies around $100$Hz.}
\end{figure*}

\begin{figure}
\epsscale{1.2}
\plotone{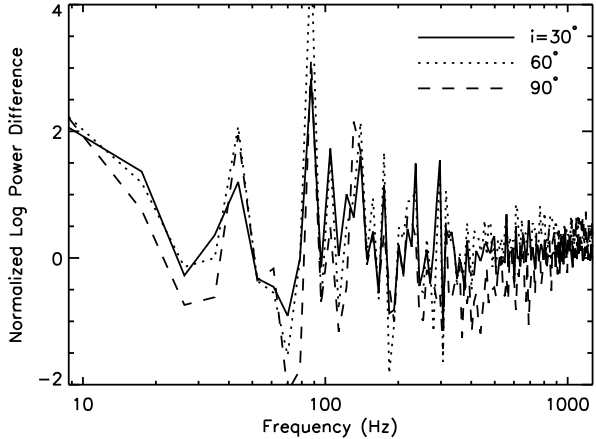}
\caption{\label{relpower}Difference in the logarithms of median 90h and 915h power spectra, normalized to their combined standard deviations for $i=30^\circ$, $60^\circ$, $90^\circ$. The median 90h power spectra are shifted to account for their lower mean power.}
\end{figure}

\begin{figure}
\epsscale{1.2}
\plotone{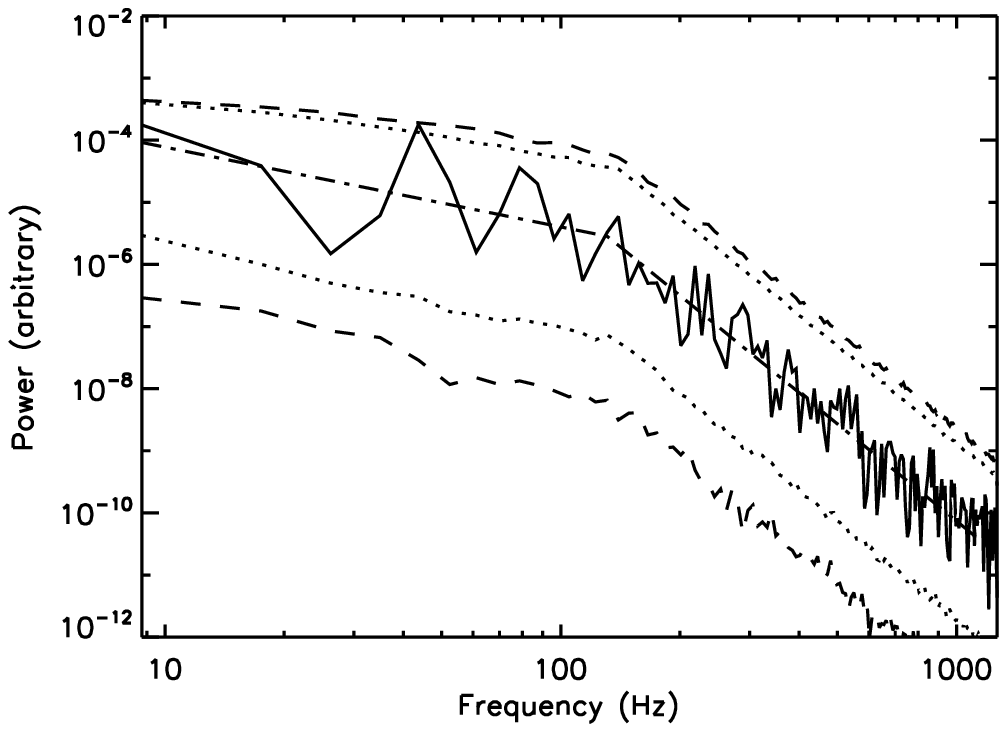}
\caption{\label{fitexample}Sample power spectrum (solid), best fit broken power law model (dot-dashed) and upper and lower 99\% (dotted) and 99.9\% (dashed) significance contours.}
\end{figure}

\subsection{Variability}
\label{var}

X-ray timing of black hole binaries has allowed the characterization of power spectra and the detection of transient QPOs \citep[for a review, see][]{remillardmcclintock2006}. High-frequency QPOs are seen in the steep power law state (SPL), while low-frequency QPOs have been observed in both the hard state and the SPL. The geometry of the accretion flow in both these states is uncertain, and there is no reason to assume complete alignment between the accretion flow angular momentum and black hole spin axes in these states. Given the time-dependent nature of the ray tracing, we can analyze the variability of the simulated accretion flows for the simplistic emission models used here to analyze the shape of their power spectra and to look for possible QPOs.

The best time sampling of the simulations is in 90h and 915h, which are used here at 8 observer azimuths, 3 inclinations and 3 observed photon energies using the thermal emission model. Each light curve captures roughly 6 (20) orbits at $r=25$M (10M), corresponding to a total observer time, $\Delta t_{\mathrm{obs}} = 0.23 M_{10}$ s, where $M_{10}$ is the black hole mass in units of $10 M_\sun$. This duration is about $1/8$ of the total precession period for the torus in the 915h simulation. 

Figure \ref{lexample} shows sample light curves and power spectra from the thermal emission model at $10$keV for an observer inclination of $i=60^\circ$. The secular trend is removed by subtracting the linear best fit from the light curve before computing the power spectrum. 

All power spectra are well fit by broken power law models of the form:
\begin{eqnarray}
  P(\nu)&=& A \nu^{-\gamma_{1}} \hspace{43pt} \nu \le \nu_b \nonumber \\
  &=& A {\nu_{b}}^{\gamma_{2}-\gamma_{1}} {\nu}^{-\gamma_{2}} \hspace{11pt} \nu > \nu_{b},
\end{eqnarray}
where $\gamma_1$, $\gamma_2$ are power law indices and the break frequency, $\nu_b$, lies near $100$Hz $M_{10}^{-1}$ in both simulations. The tilted disk power spectra tend to flatten out at the highest sampled frequencies, $\sim$1000Hz $M_{10}^{-1}$. Figure \ref{allpsds} shows median power spectra for the three different inclination angles from each simulation. The error bars are estimated from the standard deviation in $\log \mathrm{Power}$ over observer azimuths and photon energies. At higher inclinations, the peaks in the power around $100$Hz grow, especially for the tilted simulations. This would be expected from a source of excess power in the inner radii, where the larger Doppler shifts at higher inclination would enhance the signal.

To quantitatively compare the power spectra between the two simulations, the ratio between median power spectra in untilted and tilted simulations is plotted for each inclination in Figure \ref{relpower}. The values are normalized to the combined uncertainties at each frequency. The overall plots are shifted according to the mean ratio between power spectra. 

At almost all frequencies, these ratios are within $\pm 2\sigma$, and are unlikely to be observed as significant features. However, there are a few noteworthy features near $100 M_{10}^{-1}$Hz. These are particularly interesting given the finding by \citet{henisey2009} that the tilted simulation 915h contains excess power due to trapped inertial waves at $118 M_{10}^{-1}$Hz.

To assess the significance of possible features in the PSDs, the power spectrum is fit with a broken power law model. The parameters from the best fit are used to simulate many random light curves with the same parameters, and which contain no significant features. The significance is determined by comparing the values for the power at each frequency for each model power spectrum with the distribution of random ones. An example is shown in Figure \ref{fitexample}, where a single power spectrum from the 915h simulation is shown, as well as the best fit broken power law model and upper and lower 99.9\% confidence intervals from simulating random light curves. 

No obvious QPO features show up in this analysis. In several of the 915h light curves, the feature near $50 M_{10}^{-1}$Hz shows up as 99.9\% significant. It appears at high significance in more of the light curves at high inclinations. In the 90h simulations, almost all significant features are found at very high frequencies $\sim$1000$ M_{10}^{-1}$Hz. These are spurious, caused by slight errors in the fit to the post-break slope incurred by ignoring all frequencies larger than $800 M_{10}^{-1}$Hz. Including the highest frequencies in the fit can favor models with break frequencies $\sim$500 $M_{10}^{-1}$Hz, steep initial slopes and shallow post-break slopes. This occurs due to the denser sampling of the PSD at high frequencies. Simply ignoring the highest frequencies gives better results than a variety of more complicated weighting schemes. The features near $100 M_{10}^{-1}$Hz from Figure \ref{fitexample} never show up at more than 99\% significance. In general, while the feature near $50 M_{10}^{-1}$Hz in the tilted simulations is more convincing than anything from the untilted simulations, it does not appear at high enough significance at enough observer frequencies and azimuths to be identified as a QPO.

Finally, fitting the sets of power spectra provides a general idea for the range of best fit values of the broken power law parameters. The median parameters found from the tilted and untilted simulation are listed in Table \ref{vartable}, where the quoted uncertainties are the standard deviations from light curves with different observer azimuths and frequencies. Break frequencies have the units $M_{10}^{-1}$Hz. The break in slope becomes more pronounced at higher inclination as the initial slope becomes shallower while the post-break slope becomes steeper. The post-break slope is slightly shallower in the tilted simulations, while the initial slope is more strongly dependent on inclination in the untilted case.

\begin{deluxetable}{lcccccc}
\tabletypesize{\scriptsize}
\tablecaption{Broken Power Law Fit Parameters \label{vartable}}
\tablewidth{0pt}
\tablehead{
\colhead{} & \colhead{} & \colhead{90h} & \colhead{} & \colhead{} & \colhead{915h} & \colhead{} \\
\colhead{} & \colhead{$30^\circ$} & \colhead{$60^\circ$} & \colhead{$90^\circ$} & \colhead{$30^\circ$} & \colhead{$60^\circ$} & \colhead{$90^\circ$}
}
\startdata
$\gamma_1$ & $1.6 \pm 0.6$ & $0.7 \pm 0.3$ & $0.4 \pm 0.7$ & $1.2 \pm 0.3$ & $0.9 \pm 0.3$ & $0.7 \pm 0.9$ \\
$\gamma_2$ & $3.3 \pm 0.6$ & $3.8 \pm 0.3$ & $4.0 \pm 0.7$ & $3.4 \pm 0.8$ & $3.2 \pm 1.0$ & $3.7 \pm 0.8$ \\
$\nu_b$ & $80 \pm 20$ & $90\pm5$ & $100\pm20$ & $90\pm10$ & $90\pm20$ & $100\pm30$
\enddata
\end{deluxetable}

\begin{figure}
\epsscale{1.2}
\plotone{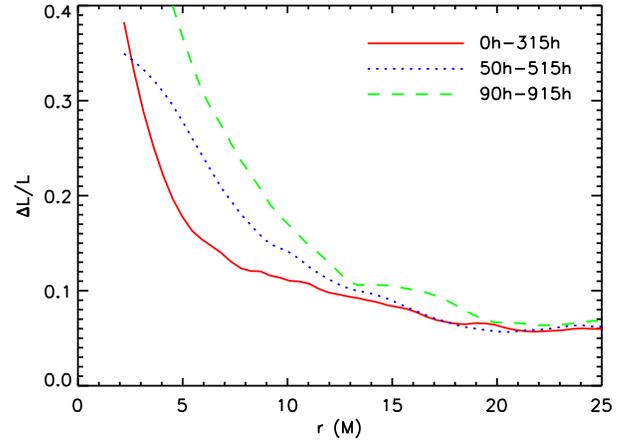}
\caption{\label{deltal}Fractional difference in shell-averaged angular momentum between tilted and untilted simulations with similar spins. The untilted simulations are nearly geodesic, while the tilted simulations are increasingly sub-geodesic with decreasing radius.}
\end{figure}

\begin{figure}
\epsscale{1.2}
\plotone{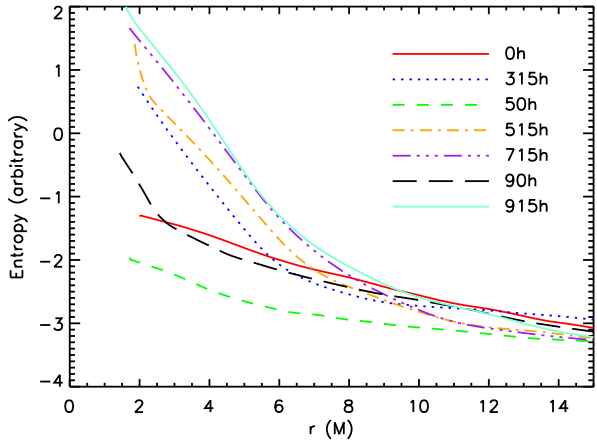}
\caption{\label{entropy}Shell-averaged entropy distributions for all simulations. Excess entropy inside $r \sim 10$M is generated by non-axisymmetric standing shocks in the tilted simulations.}
\end{figure}

\section{Physical Cause of Disk Truncation}

The observable signatures of tilted disks discussed so far are, for the most part, due to two main differences between tilted and untilted disks: tilted disks precess, and they are truncated outside $r_{\mathrm{ms}}$. \citet{fragile2007} already discussed why the simulated accretion flows precess. It is our interest to better understand the physical cause for the large truncation radius. 

The first thing to note is that rapidly rotating black holes provide more centrifugal support to an accretion disk than slowly rotating black holes. Therefore, the angular momentum extraction mechanism at play in the tilted disks must be more effective at higher spin. This is confirmed in Figure \ref{deltal}, where we plot the difference in density-weighted, shell-averaged specific angular momentum for tilted and untilted simulations of comparable spin. The angular momentum is defined as $\ell=-u_\phi/u_t$, where $u^\mu$ is the fluid four-velocity and the shell-average of a quantity $x$ is given by,

\begin{equation}
\langle x\rangle=\frac{1}{A} \int\int d\Omega \sqrt{-g} \hspace{2pt} x,
\end{equation}

where $\Omega$ is the coordinate solid angle, $g$ is the metric determinant and $A=\int\int d\Omega \sqrt{-g}$. The density-weighted shell-average of $x$ is defined as $\langle\rho x\rangle/\langle\rho\rangle$. The angular momentum profiles for the untilted simulations are nearly geodesic outside of $r \sim 5$M. Inside of $r \sim 10$M, the tilted simulations become increasingly sub-geodesic, with the higher spin cases deviating more than the lower spin ones. The same trend holds when comparing the tilted simulations to the analytic result for the angular momentum profile of material on geodesic orbits in an equatorial disk inclined $15^\circ$ to the black hole spin axis. 

\citet{fragile2008} suggested that the non-axisymmetric standing shocks that occur in the inner radii above and below the midplane of the disk may enhance the outward transport of angular momentum, causing fluid to plunge from outside the marginally stable orbit. To connect the enhanced angular momentum loss of the tilted disks with the standing shocks, we next look at a plot of the density-weighted, shell-averaged entropy profiles in Figure \ref{entropy}. Since these simulations conserve entropy except across shocks, the excess inside of $r \sim 7$M in the tilted simulations signifies the presence of extra shocks. The steepness of the entropy gradient gives some measure of the strength of these shocks. Again, we see that the effect is greatest in the simulations with the fastest spinning black holes.

\begin{figure}
\epsscale{1.2}
\plotone{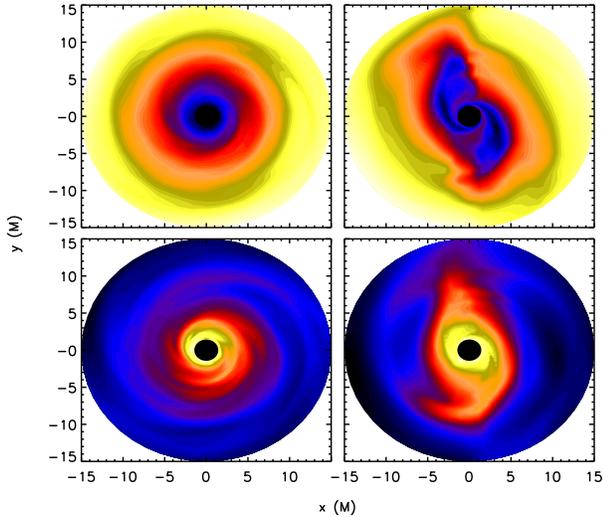}
\caption{\label{cplots}Vertically integrated contour plots of specific angular momentum (top) and entropy (bottom) for snapshots of the 90h (left) and 915h (right) simulations. The color scale is linear, increasing from blue to red to yellow to white. The non-axisymmetric shocks in 915h correspond to regions with deficit (excess) angular momentum (entropy).}
\end{figure}

Further evidence linking the sub-geodesic angular momentum profiles of the tilted simulations with the standing shocks can be found from looking at the time-dependence of the shell-averaged angular momentum. While the untilted simulation remains nearly geodesic, the tilted simulations are continuously transporting angular momentum outward from $r \sim 10$M for the first $\sim 5000$M before reaching a steady state, as would be expected from a dynamical mechanism. Finally, vertically integrated contour plots such as Figure \ref{cplots} show that the angular momentum in the tilted simulations is non-axisymmetrically distributed. The regions of depleted angular momentum correspond to the standing shocks, which appear as regions of excess entropy in the bottom panels of Figure \ref{cplots}.

\begin{figure}
\epsscale{1.2}
\plotone{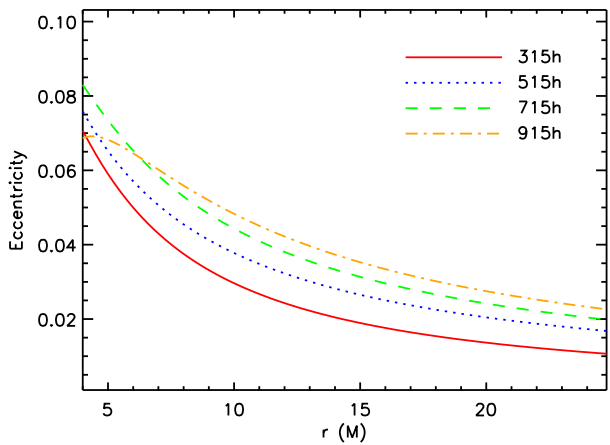}
\caption{\label{eccentricity}Shell-averaged orbital eccentricities for all tilted simulations, estimated at one scale-height in the disk. The increasing eccentricity of orbits toward smaller radii leads to a crowding of orbits at their apocenters, which, in turn, can generate standing shocks.}
\end{figure}

Following \citet{fragile2008}, we postulate that the standing shocks are caused by deviations from circular orbits near the black hole. Figure \ref{eccentricity} shows the shell-averaged eccentricities of the orbits in each simulation, estimated at one scale-height in the disk using 
\begin{equation}
e = -\frac{r}{6M}\frac{\partial (\beta \sin \gamma)}{\partial r} ~, \label{eq:ecc}
\end{equation}
where $\beta$ is the tilt and $\gamma$ is the precession of each orbital shell.\footnote{This definition of $e$ differs from \citet{ivanov1997} by a phase factor of $\pi/2$ in $\gamma$. \citet{fragile2008} used the formula from \citet{ivanov1997} without modification.} All quantities are calculated from fitting the shell-averaged disk tilt and twist (Eqs. 32 and 41 of \citealt{fragile2007}) with power laws, and using the resulting expressions in Equation (\ref{eq:ecc}). The increase in eccentricity toward smaller radii leads to a crowding of orbits near their apocenters \citep{ivanov1997}, which leads to the formation of the standing shocks. The eccentricity is larger for higher black hole spin, except inside the plunging region where the fits become poor and the eccentricity is ill-defined. Equation (\ref{eq:ecc}) may indicate how these results depend on the initial tilt of the simulations. If we assume that the strongest dependence of $e$ on tilt is through $\beta$ and that $\partial \beta/\partial r$ and $\partial \gamma/\partial r$ remain unchanged for different tilts, then equation (\ref{eq:ecc}) suggests that the eccentricity of the orbits should vary roughly linearly with the initial tilt, at least for small angles. This prediction is tentatively confirmed by a simulation we have done that started with an initial tilt of $10^\circ$.

\section{Discussion}

Tilted accretion flows will inevitably be present in a significant fraction of black hole sources with $L/L_{edd} \lesssim 0.05$ and possibly $L/L_{edd} \gtrsim 0.3$ (thick or slim disks). Using relativistic ray tracing and a set of simple emissivities, we have compared the radiation edge, emission line profiles and power spectra of simulated black hole accretion flows with a tilt of $15^\circ$ to their untilted counterparts. We find the radiation edge is independent of black hole spin, while the untilted simulations agreed with the expected qualitative trend of decreasing inner radius with increasing spin. These results for the radiation edge confirm the work of \citet{fragiletilt2009}, who used dynamical measures to locate the inner edge.

Due to the independence of inner edge on spin, the red wing of tilted accretion flow emission line profiles is also fairly independent of spin. This introduces a possible complication for attempts to measure black hole spin from sources which may be geometrically thick. In general, measurements of small spin (large inner radius) may be unreliable unless the disk is known to be untilted. A reliable estimate of a large black hole spin (small inner radius), in contrast, could rule out the presence of a tilted disk.

The blue wing can be much broader for tilted accretion flows, and the tilted-disk line profiles depend strongly on the observer azimuth as well as inclination. Since a tilted disk is expected to precess \citep{fragile2007}, highly variable emission line profiles could signify the presence of a tilted accretion flow, as pointed out by \citet{hartnollblackman2000} for warped thin disks. Since many LLAGN and X-ray binaries in the low/hard state should be tilted, time-variable emission lines should be quite common, and this effect is unlikely to significantly depend on accurate reflection spectrum modeling. Although the simulations can only be run for a short time compared to the precession time scale, precession is a possible source of low frequency quasi-periodic oscillations when the accretion flow is optically thin due to the modulation of Doppler shifts as the velocities in the accretion flow align and misalign with the observer's line of sight \citep[see][for more discussion of QPOs from precessing tilted disks]{ingram2009}.

Finally, we have studied power spectra for our simple models. We find broken power law spectra with break frequencies around $100 M_{10}^{-1}$Hz and power law indices in the range 0-2 (3-4) pre- (post-) break for both tilted and untilted simulations. Previous studies \citep{armitagereynolds2003,noblekrolik2009} found single power laws with index $\sim$2. \citet{armitagereynolds2003} found that power spectra from individual annuli are well described by broken power laws where the break frequency is close to the local orbital frequency -- the averaging of many annuli with an emissivity that falls with radius smooths the power spectrum into a single power law. We see the same behavior in our simulations; the break frequencies from power spectra of individual radial shells agree with the local orbital frequency for both simulations 90h and 915h. A break frequency $100 M_{10}^{-1}$Hz then implies a radius of $r \approx 16$M. Our broken power law spectra are therefore likely due to the fact that our emissivity peaks relatively near the outer radius used for the ray tracing, $r=25$M. A larger radial domain would likely shift the break to smaller frequencies. 

Observed break frequencies in the low/hard state are typically $\nu_b \sim 0.1-1$Hz, which may be caused by the transition from a thin disk to a thicker, ADAF flow \citep{esinetal1997}. That would imply a transition radius $r_t \simeq 200-1000 M_{10}^{2/3} M$. Our results for pre- and post-break slopes from both tilted and untilted simulations agree with those found in Cygnus X-1 \citep{revnivtsevetal2000} for an inclination $i=30^\circ$. In GRO J1655-40 \citep{remillardetal1999} our pre-break slopes agree for all inclinations. However, the PSD for that source is well described by a single power law.

There is no clear evidence in our work for high frequency QPOs due to the trapped inertial waves identified by \citet{henisey2009}, although there are more features in power spectra from the 915h simulation at higher significance than in 90h. Even when computing PSDs for sets of spherical shells from the simulations, there are no clear features in the tilted power spectra that are not also present in the untilted case. It is possible that this result could depend on the chosen emissivity. Alternatively, the excess power in trapped inertial waves could be insufficient to rise above the red noise continuum.

The independence of the inner radius of the tilted simulations on black hole spin is attributable to the extra angular momentum transport provided by the asymmetric standing shocks. These shocks are only present in the tilted simulations. Their strength scales with black hole spin, which is a necessary condition for countering the greater centrifugal support at higher spins. The standing shocks, in turn, appear to be attributable to epicyclic motion within the disk driven by pressure gradients associated with the warped structure. Again, this effect scales with the spin of the black hole, which contributes to the stronger shocks.

For small tilt angles, the orbital eccentricity scales as $e \sim \beta$. This suggests that significant deviations between the spin-dependence of the radiation edge and the marginally stable orbit should be present even at modest tilt angles $\beta \gtrsim 5^\circ$. At larger tilts, it is unclear if the increasing eccentricity will lead to an inner edge that increases with spin. This is both due to the uncertainty in the radial tilt and twist profiles $\beta(r)$ and $\gamma(r)$ at larger tilts, and to the lack of a quantitative connection between inner disk edge and eccentricity. The dynamical measures from \citet{fragiletilt2009} place the location of the inner edge in a simulation with $a=0.9M$ and $\beta=10^\circ$ closer to the location of 915h than 90h. This data point supports the idea that a noticeable departure between $r_{\mathrm{edge}}$ and $r_ {\mathrm{ms}}$ should exist between tilted and untilted disks even for $\beta \gtrsim 5^\circ$. It also suggests that at larger tilt angles, $r_{\mathrm{edge}}$ is likely to increase with spin unless the effect saturates at $\beta \approx 15^\circ$. Simulations with larger tilt angles will be able to address this question with certainty.

\begin{acknowledgements}
We thank Omer Blaes and Eric Agol for many stimulating discussions. This work was partially supported by NASA grants 05-ATP05-96 and NNX08AX59H; a graduate fellowship at the Kavli Institute for Theoretical Physics at the University of California, Santa Barbara under NSF grant PHY05-51164; and NSF grant AST08-07385.
\end{acknowledgements}

\bibliographystyle{apj}

\end{document}